\documentclass[preprint,showpacs,prc]{revtex4}

\usepackage{graphicx}
\usepackage{dcolumn}
\usepackage{bm}

\begin{document}

\title{Comment on "Fission Mass Widths in $^{19}$F + $^{232}$Th, $^{16}$O + 
$^{235,238}$ U reactions at near barrier energies" \\}

\author{T. K. Ghosh and P.Bhattacharya} 
\affiliation{Saha Institute of Nuclear Physics,  1/AF  Bidhan  Nagar,
  Kolkata  700 064, India}

\date{\today}

\begin{abstract}
A critical re-analysis of the experimental data to reject transfer fission 
component did not change the fragment mass widths and hence the conclusion 
regarding abrupt rise in mass widths with decreasing energy around Coulomb 
barrier remains unchanged.
\end{abstract}

\pacs{25.70.Jj}

\maketitle

In a recent Rapid Communication  R. Yanez et al \cite{yenez} 
reported the variations of the widths of mass distributions with 
energy contradicting the trend of increase in widths with decreasing energy 
reported by the present authors \cite{myrapid1} for the system $^{19}$F + 
$^{232}$Th and concluded that the 
difference could be due to incomplete separation of the transfer fission (TF) 
from fusion-fission (FF) by the present authors.

\begin{figure}[h]
\includegraphics*[scale=0.5,angle=-90]{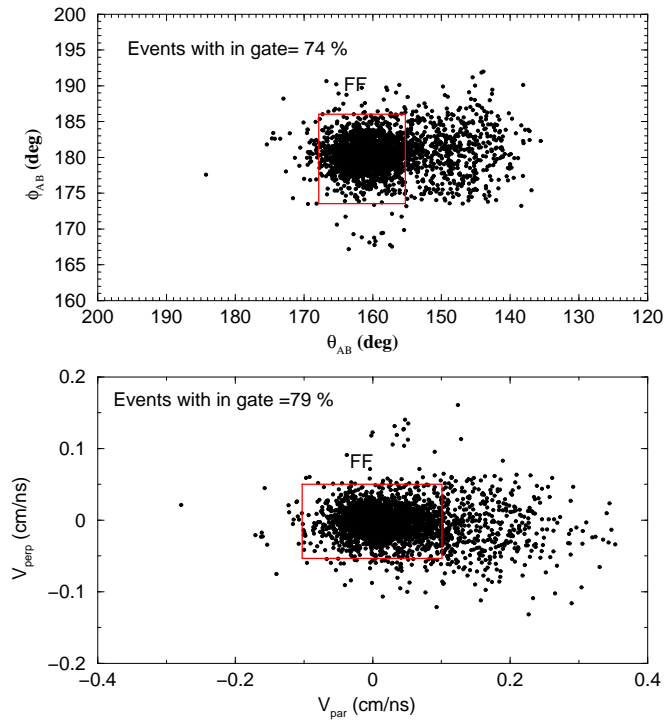}

\caption{\label{fig:fig1}~  Distributions of complementary fission fragments in 
$\theta$-$\phi$ (upper panel) and $V_{par}$-$V_{perp}$ (lower panel). The 
contour represents the gate used to select the fusion fission events. }
 
\end{figure}

We had employed the separation of  FF  from the distribution of events on 
polar ($\theta$) and azimuthal angles ($\phi$). In a re-analysis of the 
experimental data  at a center of mass energy of 85.3 MeV (shown in 
Fig.~\ref{fig:fig1}), both the procedures do not show any significant 
difference of the contours of the FF events (shown by rectangles). In 
table~\ref{tab:table1}, we have shown the events used to calculate the 
variance of the mass distributions using both gates in angles and 
velocities of the fissioning system. In fact, 
in all energies, the coincidance gates on $\theta$ - $\phi$ are more compact, 
excluding more TF events than in the correlation of velocities of the 
fissioning system. 

\begin{figure}[h]
\includegraphics*[scale=0.5,angle=-90]{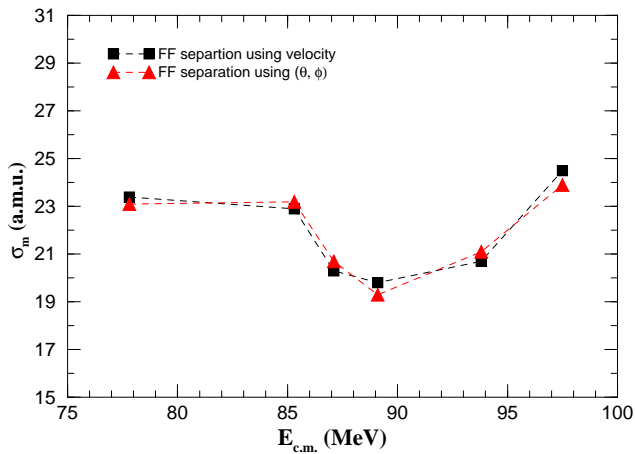}

\caption{\label{fig:fig2}~ Variance of the mass distribution for the fission 
fragments following fusion fission as function of beam energy for the 
$^{19}$F + $^{232}$Th reaction. }
\end{figure}

It may also be noted that we had used a much smaller azimuthal 
($\sim $ $12^\circ $) coverage in our detectors, hence the events from 
TF with large out of plane velocities are rejected by detector geometry. The 
corresponding mass widths 
are  also shown, and no large differences from our earlier analysis 
\cite{myrapid1} are observed which can explain the 
observed variation of about 4 a.m.u. in mass widths  as shown in 
fig.~\ref{fig:fig2}.

It may be noted that we had used long flight paths of about 52 cm and 
used {\it difference } 
of time of flights to eliminate systematic worsening of mass resolutions due to 
pulse beam time structure and variations of machine time delays with energies 
to obtain a good mass 
resolution of about 3 a.m.u \cite{myNIM}. Effects of multiple scattering in 
target on  mass widths are also  reduced. Use of {\it ratios} of fragment 
velocities to find masses considerably worsens the mass resolution even in 
our experimental arrangement to about 11 a.m.u 
\cite{myNIM} , and can be presumed 
to be considerably worsened in small flight paths of 18 cm used by 
Yanez {\it et al.,} \cite{yenez}. This is a significant difference between 
the two experimental procedure and 
 may lead to washing out the small difference in mass 
widths for both the thick and thin targets as reported in \cite{yenez}.    

\begin{table}
\caption{\label{tab:table1}~ Gated events used to calculate the mass variance}
\begin{ruledtabular}
\begin{tabular}{lcccr}
$E_{cm}$ & Events with in gate $\theta,\phi$ & Events with in gate $V_{par},V_{perp}$ \\
MeV & $\%$ &$\%$  \\
\hline
97.5 & 87 & 88 \\
93.8 & 81 & 86 \\
89.1 & 77 & 85 \\
87.1 & 75 & 84 \\
85.3 & 74 & 79 \\
77.8 & 42 & 49 \\
\end{tabular}
\end{ruledtabular}
\end{table}

\end{document}